\documentclass[a4paper,11pt]{article}
\pdfoutput=1 

\usepackage{jheppub} 

\usepackage[T1]{fontenc} 

\usepackage{mymacros}
\usepackage{lipsum}
\usepackage{multicol}
\usepackage[utf8]{inputenc}
\usepackage{subcaption}
\usepackage{lineno}
\usepackage{calc} 

\title{\boldmath Studies of Dimension-Six EFT effects in Vector Boson Scattering}

\author[]{Raquel Gomez-Ambrosio}
\affiliation[]{Institute for Particle Physics Phenomenology, Durham University, Durham DH1 3LE, UK}

\emailAdd{raquel.gomez-ambrosio@durham.ac.uk}

\abstract{
We discuss the implications of dimension-six operators of the Effective Field Theory (EFT) framework in the study of Vector Boson Scattering (VBS) in the $pp \to Z Z j j $ channel. We show that operators of dimension-six should not be neglected in favour of those of dimension-eight. We observe that this process is very sensitive to some of the operators commonly fit using LEP and Higgs data, and that it can be used to improve the bounds on the former. Further we show that other operators than the ones generating anomalous triple and quartic gauge couplings (aTGCs/aQCGs) can have a non-negligible impact on the total and differential rates and their shapes. For this reason, a correct interpretation of the experimental results can only be achieved by including all the relevant, bosonic and fermionic operators; we finally discuss how such an interpretation of experimental measurements can be done. 
}

\begin{document}
\maketitle
\flushbottom

\section{Introduction}\label{sec:intro}

Effective Field Theories (EFT) have become extremely popular in the last few years, 
\cite{Contino:2013kra,Azatov:2014jga,Contino:2014aaa,
Berthier:2015oma,Trott:2014dma,Alonso:2013hga,Jenkins:2013wua,Jenkins:2013sda,Jenkins:2013zja,Jenkins:2013fya, Artoisenet:2013puc,Alloul:2013naa, Ellis:2014dva, Falkowski:2014tna,Low:2009di,Degrande:2012wf,Chen:2013kfa,Grober:2015cwa,Englert:2015bwa,Englert:2015zra,Englert:2014uua, Biekoetter:2014jwa,Gupta:2014rxa,Elias-Miro:2013mua,Elias-Miro:2013gya,Pomarol:2013zra,Masso:2014xra, deBlas:2014mba, Corbett:2015mqf, Corbett:2015ksa, Contino:2016jqw,AguilarSaavedra:2018nen,Chala:2017sjk,deBlas:2017xtg,deBlas:2018tjm}, proving to be a robust tool for New Physics (NP) searches and BSM physics studies. In general, one can use the EFT to find the low-energy behaviour of a given UV theory (see for example Refs.\cite{Henning:2014wua,Boggia:2016asg, delAguila:2016zcb}). Alternatively, one can use EFT in an almost model-independent way: as a generalised SM extension that can be used to parametrise small deviations observed on experimental measurements of SM observables. The latter is known as the ``bottom-up'' approach and will be the one used in this work.
The main underlying idea to the bottom-up approach is to add higher dimensional operators to the (dimension-four) Standard Model Lagrangian, in a way that is consistent with the known symmetries: $SU(2) \times SU(3) \times U(1)$. Some relevant early works in this direction can be found in Refs.\cite{Appelquist:1974tg,Arzt:1993gz,Arzt:1994gp,Georgi:1994qn, Chanowitz:1987vj,Einhorn:2013kja} and some interesting reviews on the topic are Refs. \cite{Brivio:2017vri, Boggia:2017hyq, Manohar:2018aog}.
By adding new higher-dimensional terms to the SM Lagrangian, it is possible to parametrise small deviations from the original SM prediction. If such small deviations are found in the experimental data, it is possible to start \emph{mapping} the space of ``New Physics directions'', ruling out some and focusing on others.  
While few questions remain unanswered in the current picture of fundamental interactions, some seem to be quite far from an answer: How does gravity relate with the other interactions? what are dark matter and dark energy? Other important questions, however, can be tackled at LHC and from an SM perspective. The most important ones concerning the origin of electroweak symmetry breaking (EWSB).
The Higgs mechanism \cite{Higgs:1964pj,Englert:1964et,Guralnik:1964eu, Higgs:1966ev, Kibble:1967sv} is a very good description of the EWSB, but some details of the latter are still unknown to us: for example, the fact that the spontaneous symmetry breaking can be realised in a linear as well as a non-linear representation. The answer to this enigma may lay in the gauge couplings, which have only been partially studied at LEP: only some triple gauge couplings have been observed, in a very concrete energy regime and under a set of assumptions regarding the final-state radiation whereas the interactions between four gauge bosons will only be observed at LHC.
To address the last question, the detailed study of the Vector Boson Scattering (VBS) process is fundamental. In this process is characterised at tree-level by the exchange of weak gauge bosons between two quarks or a quark and an anti-quark. This means, that it gives us direct access to both triple and quartic gauge couplings. Since this is a purely electroweak process, it has a relatively small cross-section at LHC where the large QCD backgrounds dominate everywhere. However, the family of VBS processes has very particular experimental signatures: two very energetic forward jets with a big rapidity gap between them.
In this work we combine these two interesting fields, by applying EFT techniques to the study of VBS at LHC.
In particular, we study the effects of different dimension-six (\dimsix) operators in the total cross section and differential distributions of the purely electroweak contribution to the process $p p \to Z Z j j$, more commonly known as VBS(ZZ). We perform here a numerical study of the aforementioned quantities, a full analytical description will follow in an upcoming publication. 
The outline of the paper is as follows: In section \ref{sec:notation} we introduce our notation and conventions as well and define of the family of VBS processes. In section \ref{sec:anomalous} we summarise the state of the art, both on theoretical and experimental aspects, and we discuss the issue of anomalous couplings. In section \ref{sec:LEP} we compare some of the published results for dimension-six operator fits with our predictions for the cross-section of this process. In sections \ref{sec:EFTfortheGaugecouplings} and \ref{sec:warsawVBS} we study the impact of different operators of the Warsaw basis on the differential distributions. At first, we focus only on TGC/QGC operators. The full analysis is shown in section \ref{sec:EFTforTheFullProcess}, where we take into account all the Warsaw basis operators, bosonic and fermionic. 
In section \ref{sec:dijet} we focus on the main signatures for VBS: dijet observables. Showing that the effects of certain operators (in particular the four-quark ones) can be enhanced on such observables.
As anticipated, the family of VBS processes have a relatively small cross section at LHC. The situation for the $ZZ$ final state is particularly dramatic, where the QCD background is very large, with a signal to background ratio of up to $1/20$ in some phase space regions. For this reason, a rigorous treatment of the process demands also the study of the EFT effects on the corresponding background, which we perform in section \ref{sec:SignalAndBackground}.
To finalise, in section \ref{sec:generalStrategy}, we discuss a possible strategy for a global analysis including all the dimension-six operators relevant to this process. %

\section{SMEFT: notations and conventions}	\label{sec:notation}

In this work, the bottom-up approach to EFT is used.  The SM Lagrangian is extended with higher dimensional operators, consistent with the known SM symmetries. Further, we assume a linear representation for the physical Higgs field, in the form an $SU(2)$ doublet. Such a theory is commonly known as SMEFT: 

\begin{equation}
\Lag_{SMEFT} = \Lag_{SM} + \frac{c^{(5)}}{\Lambda} \op^{(5)} + \frac{1}{\Lambda^2} \sum_i c_i^{(6)} \op^{(6)}_i + \sum_{j} \sum_k  \frac{1}{\Lambda^{2+k}} c_j^{(6+k)} \op_j^{(6+k)}   .
\end{equation}
At dimension-five, $\mathrm{dim}=5$, there is only one possible operator, from Ref.\cite{Weinberg:1979sa}, which doesn't enter the process studied in this work. At \dimsix , the complete basis has 59 operators in the flavour universal case and 2499 in the most general one. In this work we will use a parametrisation of the former commonly known as the Warsaw basis, from Ref. \cite{Grzadkowski:2010es}. 

A general method to construct higher dimensional bases using Hilbert series was proposed in Ref. \cite{Henning:2015alf}. In the context of VBS, some subsets of \dimeight operators affecting quartic gauge couplings have been proposed in Refs. \cite{Eboli:2006wa,Eboli:2016kko}

\paragraph{Other EFT bases} There are additional dimension-six bases, other than the Warsaw basis. It is quite common to use the SILH basis, from Ref.\cite{Contino:2013kra} in Higgs phenomenology, however it is not optimised for multiboson processes. Instead, there is a VBS-dedicated basis, typically known as the HISZ basis, from Ref.\cite{Hagiwara:1993ck}.

\paragraph{Parameter Shifts} Adding higher dimensional terms to the SM Lagrangian has three consequences: firstly, new vertices appear. For example those with four-fermions. Secondly, the SM vertices get modified with an additional EFT contribution of the form: $V_{SM} = a \cdot g + b \cdot g \cdot c_i/\Lambda^2$, where $g$ is the SM coupling and $c_i$ is the Wilson coefficient associated with the $i^{th}$ \dimsix operator. Thirdly, there are shifts on the other SM parameters. Namely the masses, \emph{vev}, weak mixing angle and gauge fixing parameters. For a detailed discussion on the parameter shifts and gauge fixing in SMEFT see Refs.\cite{Alonso:2013hga,Ghezzi:2015vva,Helset:2018fgq,Dedes:2017zog}. 

The easiest example to understand parameter-shifts is that of the Higgs field: if we add the Warsaw basis operators to the SM Lagrangian, the Higgs part of the Lagrangian gets modified as: 

\begin{equation}\label{eq:lhiggs}
\Lag_{Higgs,EFT} = \partial_{\mu} \Phi^\dagger \partial^{\mu}  \Phi - \lambda \left( \Phi^{\dagger} \Phi - \frac{v^2}{2}  \right) + \frac{c_H}{\Lambda^2} \op_H + \frac{c_{H\Box}}{\Lambda^2} \op_{H \Box} + \frac{c_{HD}}{\Lambda^2} \op_{HD} ,
\end{equation}
where: 
\begin{multicols}{2}
\begin{itemize}
\item $\op_H = (\Phi^\dagger \Phi)^3$,
\item $ \op_{HD} = (\Phi^\dagger D_{\mu} \Phi)^* (\Phi^\dagger D^{\mu} \Phi)$,
\item $\op_{H\Box} = (\Phi^\dagger \Phi) \Box (\Phi^\dagger \Phi)$,
\end{itemize}
\end{multicols}
\noindent
and $\Phi$ is the Higgs doublet, 
\begin{equation}
\Phi = \frac{1}{\sqrt{2}} \left( \begin{array}{c}
h + v + i \phi^0 \\
\sqrt{2} i \phi^-
\end{array} \right)  \, \qquad \Rightarrow \qquad
\Phi = \frac{1}{\sqrt{2}} \left( \begin{array}{c}
h + v  \\
0
\end{array} \right),
\end{equation}
in the Feynman and unitary gauges respectively. Expanding Equation \eqref{eq:lhiggs}, the kinetic term gets modified, as well as the potential (i.e. its minimum gets shifted). To restore the correct vacuum expectation value and canonical normalization of the kinetic term, the corresponding SM parameter and field have to be redefined as: 

\begin{equation}
v \to v \left(1 + \Delta_6 (v) \right) , \qquad h \to h \left(1 + \Delta_6 (h) \right), \qquad \phi^0 \to \phi^0 \left(1 + \Delta_6 (\phi^0) \right).
\end{equation}
After some simple algebra one finds: 
\begin{equation} \label{eq:fieldredef}
v \to v \left(1 + \frac{3 v^2}{8 \lambda} \frac{c_H}{\Lambda^2} \right) , \,  
h \to h \left(1 + \frac{v^2}{\Lambda^2} c_{H \Box} - \frac{v^2}{\Lambda^2} \frac{c_{HD}}{4}  \right), \,
\phi^0 \to \phi^0 \left(1 - \frac{v^2}{\Lambda^2} \frac{c_{HD}}{4}  \right).
\end{equation}
As is customary in the literature, we re-define the Wilson coefficients as: $ \bar{c}_i = \frac{v^2}{\Lambda^2} c_i$.

The consequence of the field and parameter redefinitions in Equation\eqref{eq:fieldredef} is that each and every vertex containing the Higgs field will be dependent on the Wilson coefficients $ \lbrace c_{H \Box} , c_{HD} \rbrace $. This effect is nothing else than a wave-function renormalization in the more classical sense. 

The same phenomenon occurs for all the other fields and parameters. The case of the EW sector has to be handled with great care, specially when working on the Feynman gauge, since new linear transitions between gauge and Goldstone bosons appear. The gauge piece of the Lagrangian gets modified as: 
\begin{equation}
\Lag_{Gauge, SMEFT} = \Lag_{Gauge, SM} + 
\frac{c_{HB}}{\Lambda^2} \op_{HB} + 
\frac{c_{HW}}{\Lambda^2} \op_{HW} +
\frac{c_{HWB}}{\Lambda^2} \op_{HWB} +
\frac{c_{HG}}{\Lambda^2} \op_{HG}  .
\end{equation}

As a consequence, the gauge fields get shifted in the same fashion as shown for the Higgs field. The neutral gauge boson for example\footnote{There are many different ways to redefine the fields, depending on the gauge-fixing procedure chosen \cite{Ghezzi:2015vva, Helset:2018fgq, Dedes:2017zog}, and the treatment of the $Z-A$ transition. Eq. \eqref{eq:Zshift} represents only one option.}:
\begin{equation} \label{eq:Zshift}
Z_{\mu} \to Z_{\mu} \left( 1 + \frac{v^2}{\Lambda^2} \lbrace \,  \sin (\theta_w)^2 C_{HB} + \cos(\theta_w)^2 + C_{HWB}  \, \sin(\theta_w) \cdot \cos(\theta_w) \rbrace \right) ,
\end{equation}
where $\theta_w$ is the weak mixing angle. 
For this reason, when one studies a concrete process, it is important to also take into account these shifts and not only the EFT effects on single vertices. For example, the operator $\op_{HB}$ does not directly modify triple or quartic gauge vertices, but it enters the $Z$ field normalization and hence, any vertex containing the former.

Additionally to the field shifts, also get shifted the parameters: $\lbrace \MW, \MZ$, $\theta_w,$ $ g,  G_F \rbrace$. It is well known that these parameters are not all independent. The relation $\MZ =  \frac{\MW}{\cos(\theta_w)}$, is valid at leading-order in the SM, but has to be corrected at NLO. The same holds for the SMEFT, where this and similar relations need to be re-examined (e.g $\theta_w (g, g')$, ${\MH} ({\GF}, v)$). For this reason, the input parameter set chosen for a calculation will involve different EFT parameters depending on the choice.

\paragraph{Importance of the Input Parameter Set (IPS)}

It is important to recall that different IPS lead to different predictions, already at tree level.  While the  ``$\alpha$-scheme'': $\{ \MZ , \alpha , \GF \}$ might be convenient at  lower energies, or when using experimental measurements of EWPD,  for the energy scale of our process ($E \approx 2 \MZ$) the ``$\MW$-scheme'': $\{ \MZ , \MW , \GF \}$ is a better choice.  The modifications to the \emph{vev} are shown in Equation\eqref{eq:fieldredef}, and analogously, $\GF$ gets modified by the $\lbrace c_{\ell \ell}, c_{Hl}^{(3)} \rbrace$ operators. The one-loop renormalization of $\GF$ in SMEFT has very recently been calculated in Ref. \cite{Dawson:2018pyl}.

The Wilson coefficients of the effective theory are not observable quantities, and the $\overline{MS}$ renormalization scheme is adopted. The SM masses and the electroweak coupling, on the other hand, can be related to experimental quantities and the on-shell renormalization scheme can be adopted. 

\subsection{SMEFT amplitudes and cross-sections} \label{sec:amplitudes}

\begin{figure}[h!]
\begin{center}
\def\svgscale{0.7}
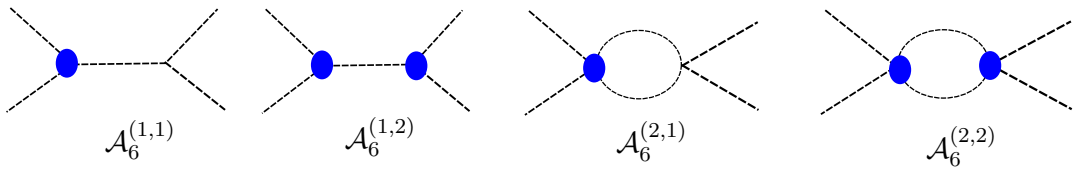 
\caption{Perturbative expansion of the EFT \dimsix amplitudes, illustrating eq.\eqref{eq:longEFTamplitude}.} 
\label{fig:amplitudes}
\end{center}
\end{figure}

In order to derive physical quantities, it is convenient to look further than the EFT Lagrangian, to $S-$Matrix elements, since those are the gauge invariant objects that can be projected into experimental quantities.  The most general EFT amplitude can be expressed as:

\begin{equation} \label{eq:longEFTamplitude}
\ampEFT = \amp_{SM} + \bar{g} \amp_6^{(1,1)}  + \bar{g}^2  \amp_6^{(1,2)}   + \frac{\bar{g}}{(4 \pi)^2} \amp_6^{(2,1)} +  \frac{\bar{g}^2}{(4 \pi)^2} \amp_6^{(2,2)}  + \frac{\bar{g}}{\Lambda^2} \amp_8^{(1,1)} + \dots   
\end{equation}
where $\bar{g} = g/\Lambda^2$. The first EFT term corresponds one \dimsix insertion in the original tree-level diagram, the second term represents two \dimsix insertions, then one insertion and one loop, two insertions and one loop, then one \dimeight \, insertion at tree level. And so on, as illustrated in figure \ref{fig:amplitudes}. 

The leading order EFT corresponds to the first term, $\amp_6^{(1,1)}$. The term following that, $\amp_6^{(1,2)}$, with two insertions at tree level, is tricky, since it is of order $(1/\Lambda^2)^2$ and the EFT expansion is only renormalizable order by order in the expansion parameter $1/\Lambda$. For this reason it is normally not included in the EFT nor NLO-EFT calculations, unless the corresponding counterterms are available. The term $\amp_6^{(2,1)}$, accounts for the NLO-EFT  amplitude, with one-loop-one-insertion diagrams.  

The perturbative EFT expansion, then, grows in different directions:  higher dimensional operators or higher loops.  If the tree level \dimeight  correction is larger than the NLO-\dimsix term will depend on the energy of the next new physics scale $\Lambda$ which, up to now, is unknown. In this work, the leading order for an EFT amplitude is defined as: $\amp = \ampSM + g_6 \amp_6^{(1,1)}$. Still, some ambiguities appear when squaring the amplitude in \eqref{eq:longEFTamplitude},
\begin{align} \label{eq:quadraticEFT}
\vert \ampEFT \vert^2 =  | \ampSM |^2 +  & \bar{g}  \, \vert \ampSM \cdot  \amp_6^{(1,1)} \vert  
+ \bar{g}^2 \, | \amp_6^{(1,1)}|^2  + \nonumber \\ & \bar{g}^2 \, | \ampSM \cdot  \amp_6^{(1,2)} |  + \frac{\bar{g}}{\Lambda^2}  \, | \ampSM \cdot  \amp_8^{(1,1)} | + \dots
\end{align}
The term $ \sim \bar{g}^2 \vert \amp_6^{(1,1)} \vert^2 $, is commonly called ``quadratic EFT'' in the literature. It can be used as an estimate for the theoretical uncertainty; it is interesting to study it in depth too, since it is positive definite and hence it can have relevant effects on the differential distributions and on the unitarity bounds of the EFT expansion. This term will be studied in the context of VBS in section \ref{sec:quadratic}.

\subsection{Off-shell effects}

Higher dimensional operators are always suppressed by a power of the new physics cut-off, $\Lambda$. This means that the expansion parameter in the perturbative EFT expansion is $E/\Lambda$, where $E$ is the energy scale of the process under study\footnote{This interpretation is consistent with the fact that the EFT coefficients $c_i$, run with the energy scale, just as every other Lagrangian parameter.}. I.e. near the $Z$-pole we can think of the EFT expansion in terms of $\MZ^2/\Lambda^2$ whereas away from the peak the EFT effects are more accurately parametrised by $\pt^2/\Lambda^2$. For this reason, the high-energy regions (tails of the $\pt$ distributions) are the ones where the EFT effects are expected to be largest (i.e. $ E_1 / \Lambda >> E_2/\Lambda $, for $E_2$ on the pole and $E_1$ on the tail).

The VBS process is defined by two very energetic jets. This means that the $\pt (j)$ and $m_{jj}$ distributions are privileged kinematic variables where to expect EFT effects.  

\subsection{VBS: Definition of the process} \label{sec:VBSdef}

The family of vector boson scattering processes is very interesting, since it lays at the heart of electroweak symmetry breaking. Some works describing the details of these processes are Refs.\cite{Accomando:2006mc,Rauch:2016pai,Biedermann:2016yds, Baur:1990mr, Ballestrero:2011pe}. Unitarity and gauge-invariance are conserved in this process thanks to a series of cancellations between Feynman diagrams, and fundamentally thanks to the introduction of the Higgs boson, see Ref. \cite{Kleiss:1986xp,Philippides:1999gx}, for these reasons, the VBS represent a set of privileged channels for New Physics studies. For some applications of NP searches to VBS see for example Refs. \cite{Ballestrero:2012xa, Kilian:2014zja, Delgado:2015kxa,Dobado:2015hha, Sekulla:2016yku, Gomez-Ambrosio:2016ybh,BuarqueFranzosi:2017ugz, Brass:2018hfw, Delgado:2018ldw}.

The possible definitions of the VBS process are multiple. Typically, there are substantial differences between the theoretical definitions, in terms of initial and final states, and the experimental definitions, that constrain the phase space of the final states as well. In particular, it is common that the experimental analyses impose certain cuts to try and decouple the vector-boson fusion (VBF) process, where a Higgs boson in the s-channel is produced from the exchange of weak bosons between a quark and an antiquark. This way, the VBF channel is studied in dedicated Higgs analyses, whereas the VBS channels belong to the \emph{multiboson} analyses.

Regarding such cuts, there are different VBS-regions that are widely accepted, but they obviously lead to different results.  In the most recent LHC results, in Ref.\cite{Sirunyan:2017fvv}, the VBS(ZZ) is defined as the purely electroweak component of $p p \to Z Z j j \to \ell \bar{\ell} \, \ell' \bar{\ell'} j j  $, measured in the region defined by the following cuts:
\begin{multicols}{2}
\begin{itemize}
\item $\pt (j) > 30$ GeV
\item $ m_{jj} > 100$ GeV
\end{itemize}
\end{multicols}
\noindent One can define a VBS-enriched region, with the additional cuts:
\begin{multicols}{2}
\begin{itemize}
\item $ \Delta \eta (j_1 j_2) > 2.4$
\item $ m_{jj} > 400$ GeV
\end{itemize}
\end{multicols}
\noindent 
which was also used in some parts of that analysis. In this work we chose a compromise between both regions, and apply the cuts: 
\begin{multicols}{2}
\begin{itemize}
\item $\pt (j) > 30$ GeV
\item $ \Delta \eta (j_1 j_2) > 2.4$
\item $ m_{jj} > 100$ GeV
\end{itemize}
\end{multicols}
As anticipated, further cuts impose the two $Z$ bosons to be on-shell, to remove the VBF contamination. Nevertheless it is important to keep in mind that the experimental cut defining ``on-shell'' $Z$'s ($ M_{\ell \bar{\ell}} \in \left[ 60 , 120 \right] \GeV$ ) is not strictly the same thing as the theory definition for the $Z$ pole. 

For a matter of clarity, in this work we focus only the process  $p p \to Z Z j j $ before the on-shell decay. The difference between the process $p p \to Z Z j j$ followed by the on-shell decay: $Z \to \ell \bar{\ell} $ and the process $p p \to \ell \bar{\ell} \, \ell' \bar{\ell'} j j  $ in the aforementioned fiducial region, is negligible. In general when applying multivariate-analysis techniques the first definition is preferred, since it populates the phase space in a more effective way. 

For a rigorous EFT treatment, the same study should be performed, to make sure that the difference between both options is also negligible in SMEFT. It is clear that new operators will come into play, mainly the ones connecting quarks and leptons in the final state, however intuition and experience tell us that the VBS-cuts will most likely remove the bulk of that contribution. 
%



\section{Weak-boson triple and quartic couplings}\label{sec:anomalous}

\subsection{Anomalous couplings}
Anomalous gauge couplings were introduced in Ref. \cite{Gaemers:1978hg}, at a time when the EWSB mechanism had not been thoroughly tested yet, and before the Higgs boson was discovered. Such couplings, defined in terms of \emph{ad-hoc} variations on the Lagrangian parameters, might be good in a first approximation, but present serious theoretical inconsistencies. The main problem being that they violate gauge invariance and unitarity beyond the leading order.  

The EFT approach aims to parametrise small deviations from the SM predictions, which are currently being tested with unprecedented precision at LHC. In that regard, a more consistent approach than the anomalous couplings is needed. In particular, one that is consistent at next-to-leading order. The SMEFT approach considered in this work, in terms of dimension-six operators, represents an optimal solution to the anomalous-coupling problem: it can be understood as a SM Lagrangian where \emph{all} the parameters are anomalous, but in a way consistent with the QFT rules. 

Numerous works regarding anomalous gauge couplings in SMEFT can be found in the literature. Some or the earliest studies are Refs. \cite{Dobado:1990zh,Grinstein:1991cd, Hagiwara:1992eh,Hagiwara:1993ck,Wudka:1994ny}, and more current ones can be found for example in Refs. \cite{Fabbrichesi:2015hsa, Falkowski:2016cxu,Corbett:2014ora,Corbett:2017ecn,Mebane:2013zga,Rahaman:2017aab}. In the upcoming sections we  will investigate and discuss the differences between allowing the EFT operators only on the weak-couplings (in the spirit of the anomalous-coupling approach) or allowing them to occur anywhere. 

\subsection{SMEFT for triple and quartic gauge couplings}

As part of the legacy of the LEP experiment, it is common in the experimental analysis to study triple gauge couplings (TGCs) in multiboson production channels. For some LEP/LEP2 results, see Refs.\cite{Jousset:1998inf, delaCruz:2002zz, Schael:2004tq, Abdallah:2008sf,Schael:2009zz} and for some LHC analysis see  Refs.\cite{Aaboud:2017rwm,ATLAS:2017eyk,Sirunyan:2017zjc}. Quartic gauge couplings (QCGs), on the contrary, were more difficultly accessed by LEP/LEP2 (for example in $e^+ e^- \to \gamma \gamma \nu \bar{\nu}$ and $e^+ e^- \to \gamma \gamma q \bar{q}$ were studied in Refs.\cite{ALEPH:2001aa, Abbiendi:2004bf}) and not all QGCs where accessible.  

This fact makes the QGCs an interesting goal for the LHC experiments, where QGCs are studied in the VBS channels, for example in the analysis \cite{Sirunyan:2017fvv}. Still, it is important to emphasise that this approach is extremely misleading, since it implies identifying a collection of thousands of Feynman diagrams with a single (off-shell!) vertex, and more importantly, it implies the assumption that triple and quartic gauge couplings are not originated simultaneously through the EWSB. For this reason an analysis in terms of cross-sections, differential distributions or (pseudo-)observables should always be preferred. 

At the LEP experiment, the $s$-channel production of $Z/W$-bosons was very well under control, as well as their decays, since the electroweak radiation could be deconvoluted pretty accurately from the process. This made it posible to treat triple-gauge couplings as \emph{pseudo-observables}, which could be measured by the experiments. This is not the case any more at the LHC, and hence, one should not aim at ``measuring'' triple or quartic electroweak couplings. 

Pseudo-observables are well-defined theoretical quantities that can be measured in the experiment, a classic example is that of the set of EWPD. The most promising alternative for LHC physics relies on the study of the residues of S-Matrix poles, which are by default gauge-invariant quantities. For some work in this direction, see Refs. \cite{Bordone:2015nqa,David:2015waa, Greljo:2015sla,Brivio:2017btx,Gainer:2018qjm}, and the reviews \cite{deFlorian:2016spz,Boggia:2017hyq}.

The impact of \dimsix operators on triple gauge couplings has been studied in Refs.\cite{Corbett:2014ora, Corbett:2017ecn}. And the set of \dimeight operators affecting quartic gauge couplings has been studied in depth in Ref. \cite{Degrande:2013kka,Eboli:2016kko,Eboli:2006wa}, and other \dimeight subset, relevant for diboson studies has been shown in Ref. \cite{Bellazzini:2018paj}. Similar studies to the one presented here, tackling vector boson scattering in the \dimeight \, basis have been presented in Refs. \cite{Kalinowski:2018oxd,  Zhang:2018shp}, as well as works on VBS in the context of the electroweak chiral Lagrangian, in Refs. \cite{Delgado:2017cls,Arganda:2018ftn}. However there is no study, to our knowledge, addressing \dimsix \, effects in the context of VBS. 
\subsection{Subsets of operators and gauge invariance}

Each of  the operators in the Warsaw basis is independently gauge-invariant and in principle, it is possible in a tree-level study to select a subset of operators without breaking this gauge invariance. This situation however, will not hold beyond tree-level, where different operators enter through the \dimsix \, counterterms, and the full basis is needed for UV-renormalization. For a further discussion on the SMEFT renormalization see Refs.\cite{Ghezzi:2015vva,Jenkins:2013wua,Jenkins:2013zja,Alonso:2013hga}.

Gauge invariance is also broken if the effects of certain \dimsix or \dimeight operators are only included on a certain vertex and not in other vertices or wave function normalizations. For example, it is gauge-invariant to include only $\op_W$,$\op_{HW}$ and $\op_{HWB}$, neglecting $\op_{HB}$. But it is not completely rigorous: $\op_{HB}$ enters every vertex containing  a $Z$ field, as shown in equation \eqref{eq:Zshift}, and every expression containing the weak-mixing angle, and it might enter in different ways depending on the IPS chosen. The same holds for other operators, mainly $\op_{\ell \ell}$ and $\op_{Hl}^{(3)}$, that enter as corrections to $\GF$.  
 
%


\section{EFT for the Gauge Couplings} \label{sec:EFTfortheGaugecouplings}

To allow a straightforward comparison with the existing literature, in this section we study the impact of a handful of EFT operators. In particular, the operators that directly affect triple and quartic gauge vertices, and are CP-even.\footnote{The analysis can be easily extended to the CP-odd case, by adding $\op_{\widetilde{W}}$, $\op_{\widetilde{HW}}$ and $\op_{\widetilde{HWB}}$.}

\begin{itemize}
\item $\op_W = \epsilon^{ijk} W_{\mu}^{i \nu} W_{\nu}^{j \rho} W_{\rho}^{k \mu} $
\item $\op_{HW} = H^{\dagger} H W_{\mu \nu}^I W^{\mu \nu I}$
\item $\op_{HWB} = H^{\dagger} \tau^I H W_{\mu \nu}^I B^{\mu \nu}$
\end{itemize}
For this preliminary study we generated $3 \cdot 10^5$ events for the process defined in section \ref{sec:VBSdef} using a modified version of the \texttt{SMEFTsim} package \cite{Brivio:2017btx}, interfaced with \texttt{Madgraph5$\_$aMC@NLO} \cite{Alwall:2014hca} via \texttt{FeynRules} \cite{Alloul:2013bka} and \texttt{MadAnalysis5} \cite{Conte:2012fm}. We studied the impact of each of the three TGC/QGC operators separately, as well as the sum of them, following the definition of leading-order EFT given in section \ref{sec:amplitudes}.

\begin{figure}[h!]
\centering
\includegraphics[scale=0.35]{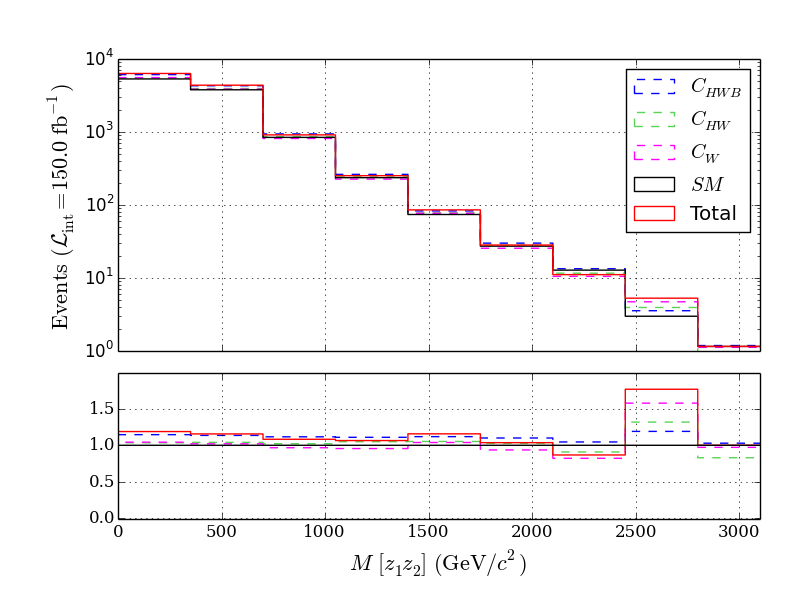}
\hfill
\includegraphics[scale=0.35]{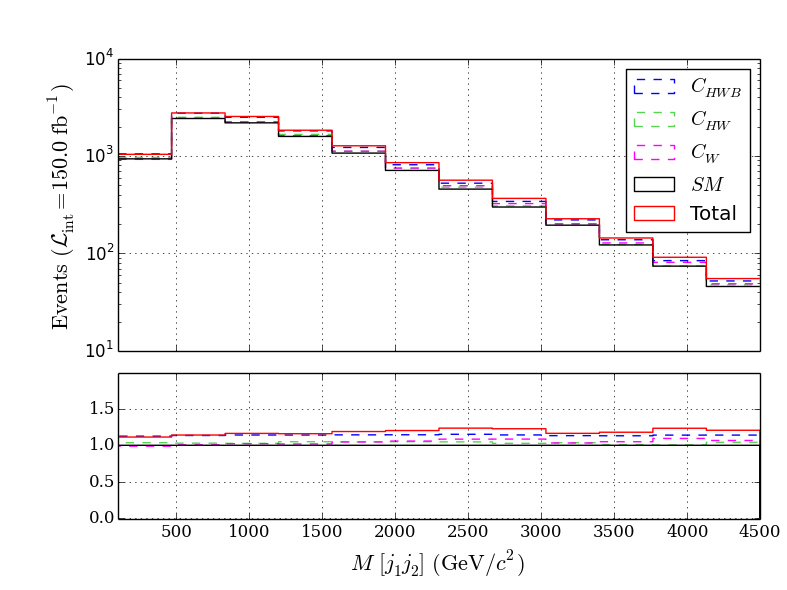}
\includegraphics[scale=0.35]{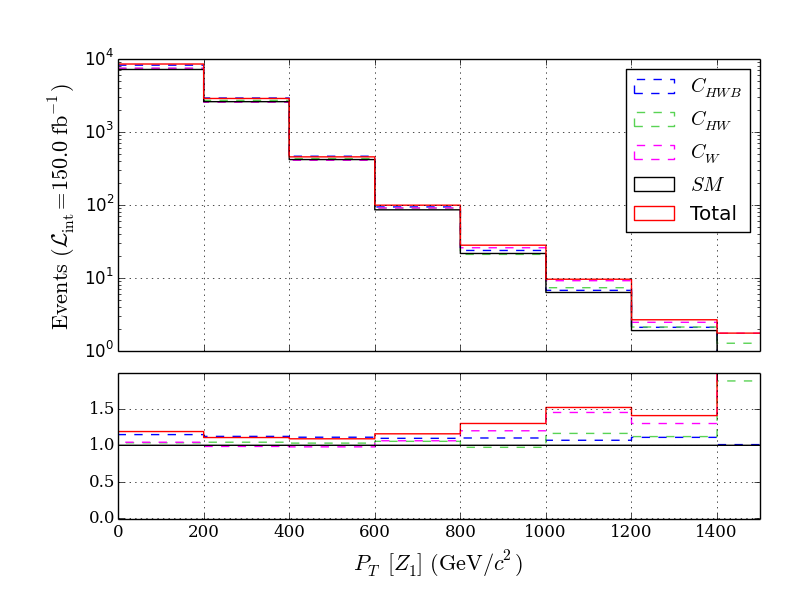}
\hfill
\includegraphics[scale=0.35]{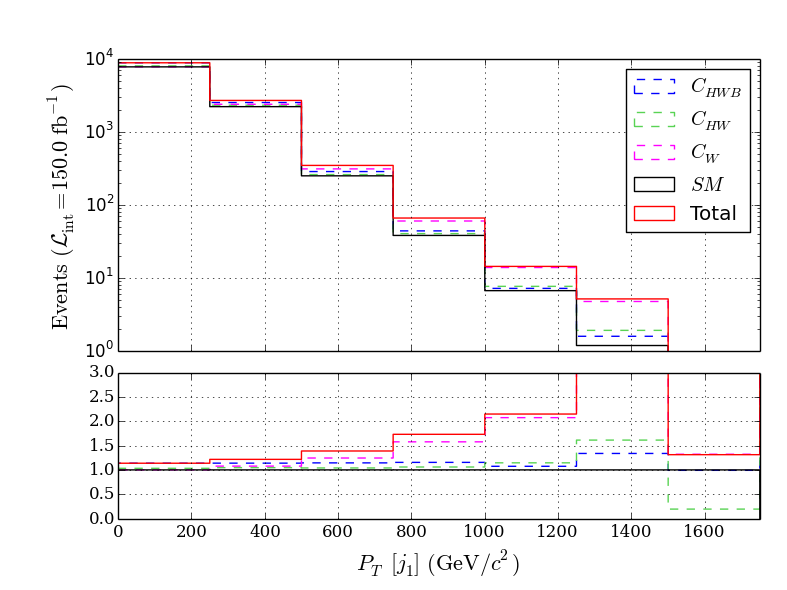}
\caption{\label{fig:relativeImpact} Here we show the impact of the 3 Warsaw basis operators that affect the triple and quartic gauge couplings. We set the values $\bar{c}= c \frac{v^2}{\Lambda^2}= 0.06$ which correspond to $c=1$ for $\Lambda=1 \rm{TeV}$. However it is important to recall that one of the main assumptions of the EFT is that there are no new \emph{light} resonances, and hence in a histogram like this one we are implicitly assuming $\Lambda > 3$TeV, while keeping $\bar{c}= c \frac{v^2}{\Lambda^2}= 0.06$. }
\end{figure}

In Figure \ref{fig:relativeImpact} we see the impact of each of the three operators individually and the sum of them, for four different observables: the invariant mass of the two final $Z$ and that of the two final jets, and the transverse momentum of the leading $Z$ and leading jet. For the numerical values of the coefficients. In this section, we choose the \emph{democratic} values 
$\bar{c}_W = \bar{c}_{HW} = \bar{c}_{HWB} = 0.06$, which correspond to ${c}_W = {c}_{HW} = {c}_{HWB} = 1$ with $\Lambda = 1$TeV.  In section \ref{sec:LEP} we will discuss the case of the available \emph{best-fit} values. 

We observe that the EFT effects on the invariant mass distributions are relatively homogeneous, in particular in the two-jet case. The VBS signature is characterised by two very energetic jets and the high energy phase space is quite well populated. On the contrary, for the $ZZ$ invariant mass we find that at very high energies we reach the
limit where the SM production tends to zero, and the EFT effects become sizeable. This effect is of course dominated by the Monte Carlo uncertainty, but it points us to a region that should be studied in detail. Such regions where the SM production becomes negligible are also those where the quadratic-EFT of equation \eqref{eq:quadraticEFT} will have a dominant role. This will be discussed in section \ref{sec:quadratic}.

The case transverse momentum distributions are very interesting themselves. We find out that the EFT effects get enhanced on the tails of such distributions, which is something that we would have expected a priori for all four observables, but is not so pronounced as expected for the invariant masses. For both cases: $p_T (Z_1)$ and $p_T (j_1)$, we also reach the regime where the SM production is negligible but the EFT effects remain.

%

\subsection{Comparison with LEP and Higgs bounds} \label{sec:LEP}

Several works have appeared in the last years, where SMEFT predictions are compared with LEP data (in Refs. \cite{Berthier:2015oma,Berthier:2016tkq,Brivio:2017bnu}) and LHC data (the SILH basis in Refs. \cite{Englert:2015hrx,Englert:2017aqb} and more recently the Warsaw Basis in Refs. \cite{deBlas:2017wmn, Ellis:2018gqa,Alioli:2018ljm}). 

In order to be consistent, a global fit of the full Warsaw basis would be desirable. For this reason it is necessary to include as many measurements as possible, including fiducial cross-sections as well as differential distributions. 

\begin{figure}[h!]
\centering
\includegraphics[scale=0.25]{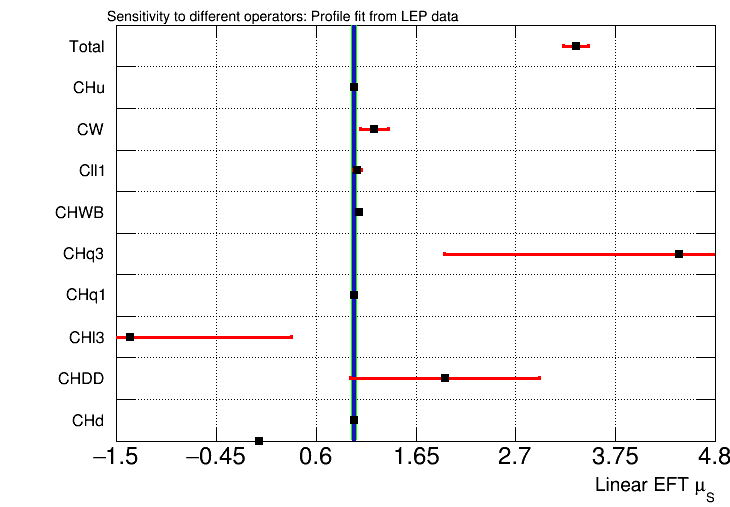}
\hspace{2cm}
\includegraphics[scale=0.25]{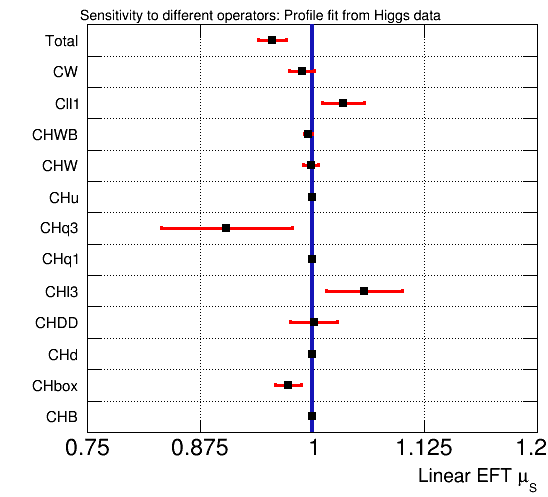}
\caption{\label{fig:root} Relative contribution given by each of the best-fit values. Operators that enter the VBS(ZZ) cross-section but have not been fitted in the past have been set to zero in these figures. On the left we use the values from the LEP fit, Ref.\cite{Brivio:2017bnu}  and on the right from the LHC fit Ref.\cite{Ellis:2018gqa} mentioned in the text. 
}
\end{figure}

In this section we study how do the published best-fit values enter the VBS(ZZ) fiducial cross-section. In Figure \ref{fig:root} we show the signal strength $ \mu = \frac{\sigma_{EFT}}{\sigma_{SM}}$ for the central values given by the profile fit of the operators (black, points) and their $95\%$ confidence level bounds (red, error bars)\footnote{The data are taken from the aforementioned papers. For the LEP fit we took the values corresponding to the $\MW$ IPS and to a $1\%$ theoretical uncertainty. For the Higgs fit we took the values in Table 4 of Ref.\cite{Ellis:2018gqa}, but it was not possible to clarify some of the details with the authors}.  

For the LEP case, we find that the values for some operators are off by a large amount ($\geq 200\%$) which means that this channel could be an interesting one to constrain such a fit better. This is not surprising since, as precise as it was, LEP is still a ``low-energy'' experiment\footnote{ An important step of any EFT calculation requires the matching of the EFT calculation, at a low energy scale $E_0$ with the ``new physics'' scale $\Lambda$, to account for the running of the EFT coupling.}, compared with the energies considered here and throughout LHC Run-2. Moreover, the treatment of the radiation in LEP raises some doubts on the applicability of such measurements for EFT searches, as discussed in section I.5 of Ref.\cite{Boggia:2017hyq}. 

In the LHC case, we find that the fit is more precise and its predictions are compatible with the ones here. The total contribution including its error bar, however, still departs from the SM expected value. Further, it is important to recall that only a subset of EFT operators is included in the former fits, and hence some directions in the phase space have not been tested. The results shown on Figure \ref{fig:root} further show that only one of the four-fermion operators ($\op_{\ell \ell}$, the one entering $\GF$) is well constrained, whereas other 11 of these operators enter the VBS cross-section. Contrary to the common belief that four-fermion operators are maximally constrained from LEP measurements, we see that, while this might be the case for most of the leptonic operators, it is not the case for four-quark ones.  


\newpage

\subsection{Linear and Quadratic contributions} \label{sec:quadratic}

Another issue that should be handled with care is the treatment of quadratic \dimsix terms. This effect was discussed in section \ref{sec:amplitudes},  and it was thoroughly studied in section I.4.7 of Ref. \cite{Boggia:2017hyq} and in Refs. \cite{Brehmer:2015rna,Biekotter:2016ecg}. The quadratic contribution to the cross-section is not included in the SMEFT predictions, as a matter of consistency: its perturbative order is higher than the linear term, equivalent to the \dimeight operators that are not included either. However, this quadratic term is, by definition, always positive, and hence can have a large impact on the behaviour of the final distributions. 
There are some situations where the quadratic contribution should be carefully studied:
\begin{itemize}
\item In the regions when the SM prediction becomes very small. In that case the interference term $SM \times EFT_6$ is dominated by the $EFT$ contribution, which is expected to be large. If this is the case, the quadratic term must also be quite large and should be calculated as part of the theoretical uncertainty. 
\item In the cases where the linear \dimeight interference term is included ($SM \times EFT_8$), since they are both of the same perturbative order, $\op (\Lambda^-4)$.   
\end{itemize}

\begin{figure}[h!]
\centering
\includegraphics[scale=0.35]{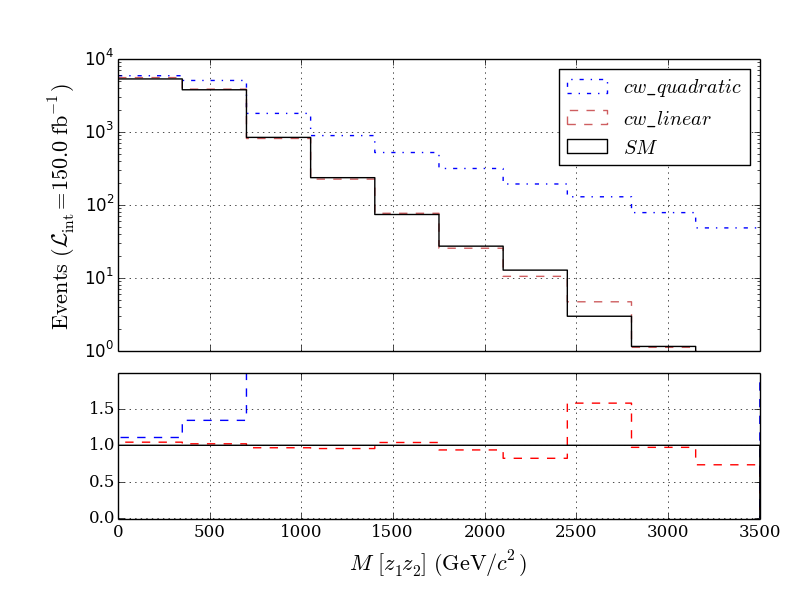}
\hfill
\includegraphics[scale=0.35]{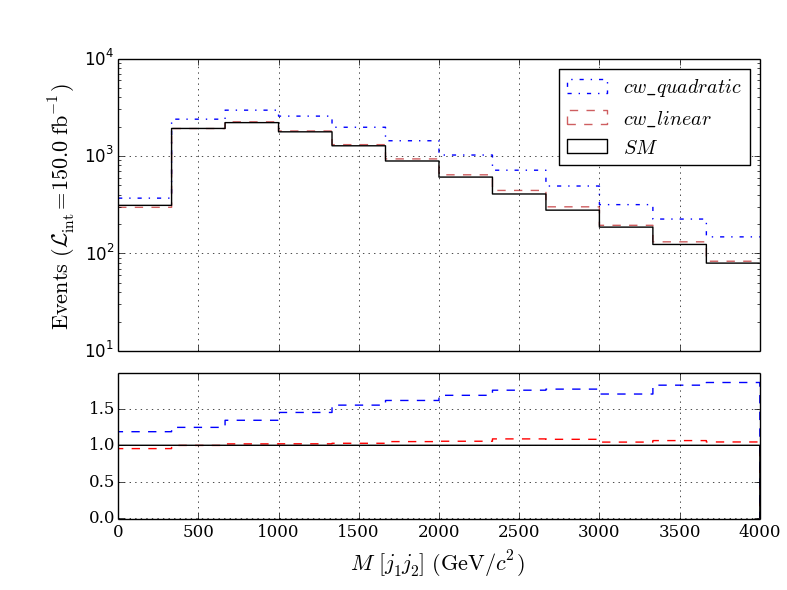}
\includegraphics[scale=0.35]{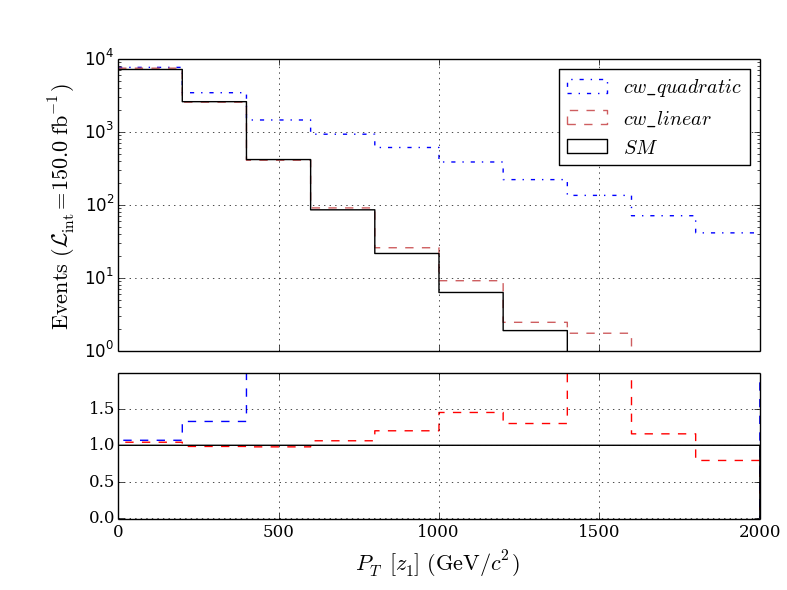}
\hfill
\includegraphics[scale=0.35]{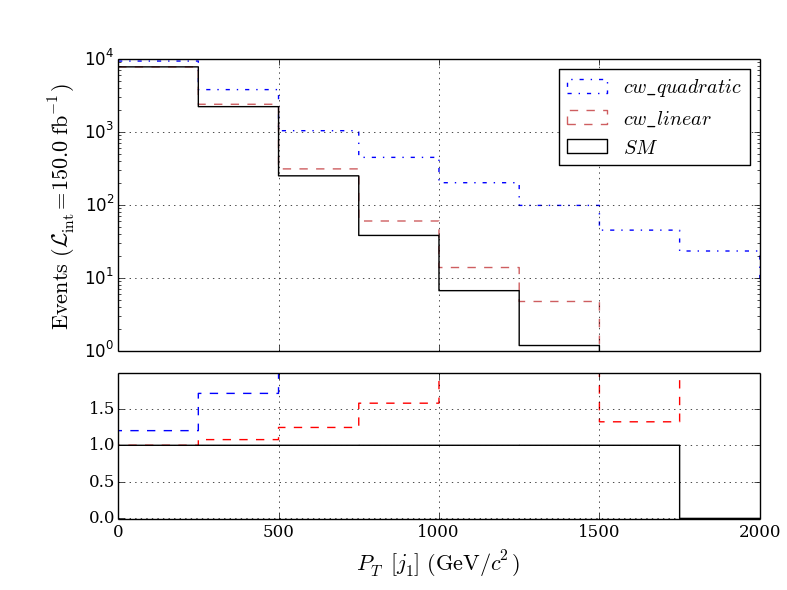}
\caption{\label{fig:quad} 
Quadratic effects on some kinematic distributions. The red lines represent the linear contribution to the cross section (\dimsix interference with the SM) and the blue lines represent the previous contribution plus the purely \dimsix term.  We take the canonical values $\bar{c}_W = 0.06$ and $\Lambda = 1 \TeV$. The quadratic effects are relevant in general on the tails of the distributions, and in particular on the bins where the SM production vanishes. They also play a very important role in the bins where the interference with the SM is negative, since they may restore unitarity. 
$\Lambda = 1 \TeV$ is the \emph{worst-case-scenario}, for higher values of $\Lambda$ the difference between the linear and quadratic terms decreases.
} 
\end{figure}

The latter situation, where \dimeight operators where compared with \dimsix quadratic terms, was recently studied in Ref. \cite{Hays:2018zze}. 
In figure \ref{fig:quad} we show an example of the impact of including the quadratic contribution in some key distributions, for the canonical value $\Lambda = 1 \TeV$. 
Some good news is that, for higher values of the New-Physics scale, $\Lambda$, the ratio between quadratic and linear contributions is expected to smoothen.  I.e. $r = \frac{(1/\Lambda^2)}{(1/\Lambda)} = \frac{1}{\Lambda}$, hence $r_2 << r_1$ for $\Lambda_2 >> \Lambda_1$.

 %

\clearpage
\section{The Warsaw Basis in VBS} \label{sec:warsawVBS}

\subsection{EFT for the full process} 
\label{sec:EFTforTheFullProcess}

\begin{figure}[h]
\centering 
\includegraphics[scale=0.325]{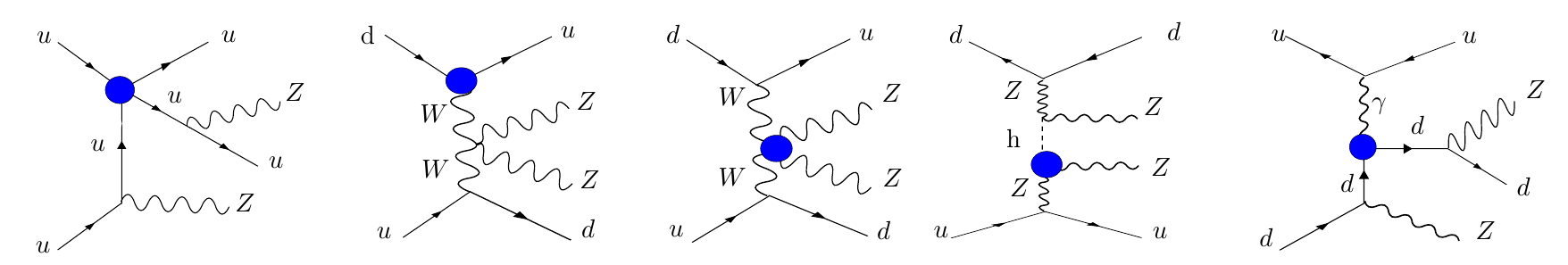}
\caption{\label{fig:signalFD} Some of the Feynman diagrams contributing to the VBS(ZZ) process in \dimsix EFT. The blobs represent the dimension-six insertions.}
\end{figure}

In this section, we investigate the effects of including all the Warsaw-basis operators on the computation of the VBS(ZZ) cross-section. Some examples of the Feynman diagrams that contribute to the process are shown in figure  \ref{fig:signalFD}.

In particular we found, numerically, the following expressions for the bosonic contribution to the total cross-section: 

\begin{equation} \label{eq:bosonicinterf}
\begin{split}
\frac{\sigma_{EFT,bosonic}}{\sigma_{SM}} \approx 
1. & 
+ 0.047 \, \, \bar{c}_{HB} 
- 0.053 \, \, \bar{c}_{ H\Box} 
- 0.0021 \, \, \bar{c}_{\widetilde{HB}}
+ 0.010 \, \,  \bar{c}_{Hd} 
- 1.84 \,  \, \bar{c}_{HD}  \\ & 
- 3.86 \,  \, \bar{c}_{Hl^{(3)}}
- 0.017 \,  \, \bar{c}_{Hq^{(1)}} 
+ 5.61 \,  \, \bar{c}_{Hq^{(3)}} 
- 0.033 \,  \, \bar{c}_{Hu} 
+ 0.59 \,  \, \bar{c}_{HW} \\ &  
- 0.0041 \,  \, \bar{c}_{\widetilde{HW}}
- 0.69 \,  \, \bar{c}_{HWB} 
- 0.022 \,  \, \bar{c}_{\widetilde{HWB}}
+ 0.23 \,  \, \bar{c}_{W}
- 0.086 \,  \, \bar{c}_{\widetilde{W}}.
\end{split}
\end{equation}
And for the fermionic contribution:
\begin{equation} \label{eq:fermionicinterf}
\begin{split}
\frac{\sigma_{EFT,fermionic}}{\sigma_{SM}} \approx  
1. & 
- 3.23 \cdot 10^{-6} \,  \, \bar{c}_{dd} 
- 2.89 \cdot 10^{-6} \,  \, \bar{c}_{dd}^{(1)} 
- 3.86  \,  \, \bar{c}_{\ell \ell}^{(1)} 
+ 0.0010 \,  \, \bar{c}_{qd}^{(1)} \\ & 
+ 1.80 \cdot 10^{-20} \,  \, \bar{c}_{qd}^{(8)} 
- 1.93 \,  \, \bar{c}_{qq}^{(1)} 
- 2.57 \,  \, \bar{c}_{qq}^{(11)} 
- 14.3 \,  \, \bar{c}_{qq}^{(3)} 
- 10.3 \,  \, \bar{c}_{qq}^{(31)} \\ & 
- 0.0049 \,  \, \bar{c}_{qu}^{(1)} 
- 2.51 \cdot 10^{-20} \,  \, \bar{c}_{qu}^{(8)}  
+ 0.00020 \,  \, \bar{c}_{ud}^{(1)} \\ &
+ 1.62 \cdot 10^{-21} \,  \, \bar{c}_{ud}^{(8)} 
- 0.0010 \,  \, \bar{c}_{uu} 
- 0.00099 \,  \, \bar{c}_{uu}^{(1)} .
\end{split}
\end{equation}
Both of this expressions have been extracted from simple numerical analysis of relatively small Monte Carlo samples and hence are dominated by the MC uncertainty. Which we estimate of the order of 10\% for each of the interference terms.  The only purpose of displaying them here is to give an impression of the relative sensitivity of this process to the different EFT operators.

It is interesting to observe that the bosonic interference is generally positive, while the fermionic one is generally negative. This means that in the case that all the Wilson coefficients would have the same sign, both interferences could extensively cancel, giving rise to a very small SM deviation in the total cross section. For this reason it is fundamental to define observables and regions where the EFT effects are maximised.

To understand the impact of the full \dimsix basis, we defined different benchmark scenarios where we study the differential distributions for the most interesting VBS observables.

\begin{figure}[h!]
\centering 
\includegraphics[scale=0.35]{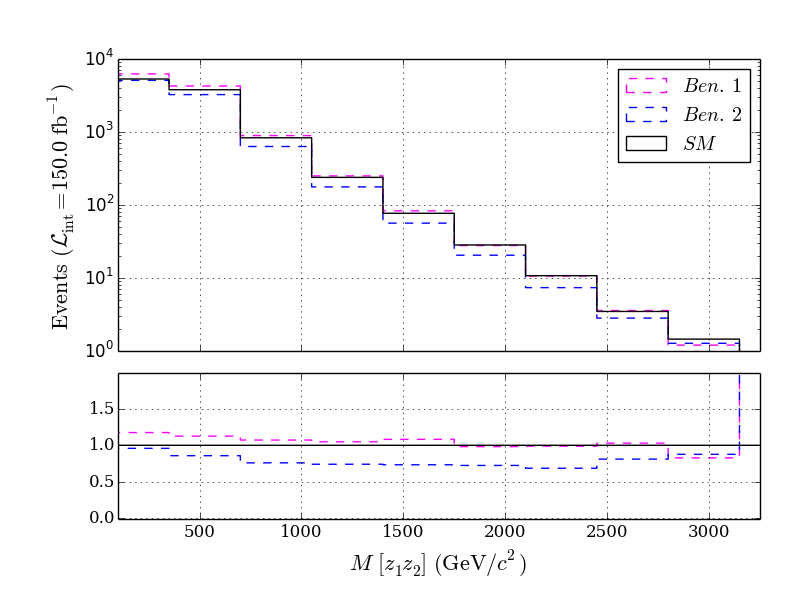}
\hfill
\includegraphics[scale=0.35]{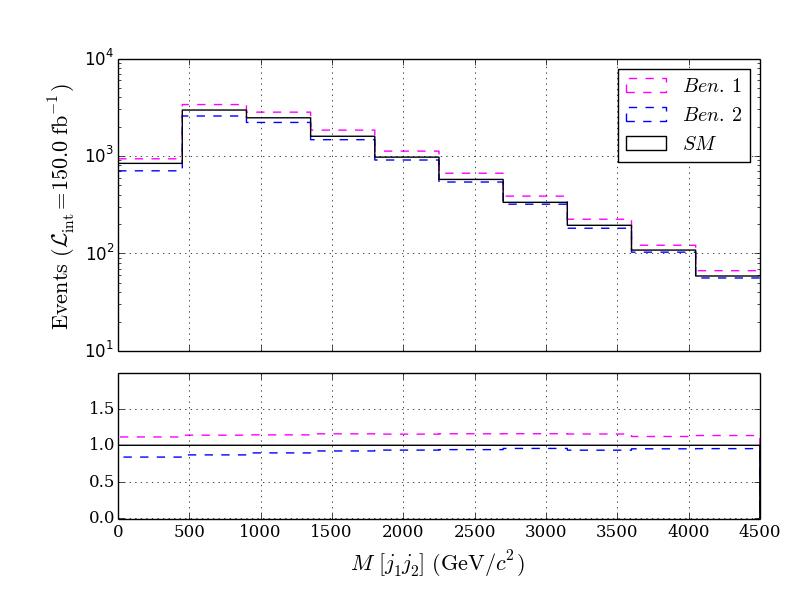}
\hfill
\includegraphics[scale=0.35]{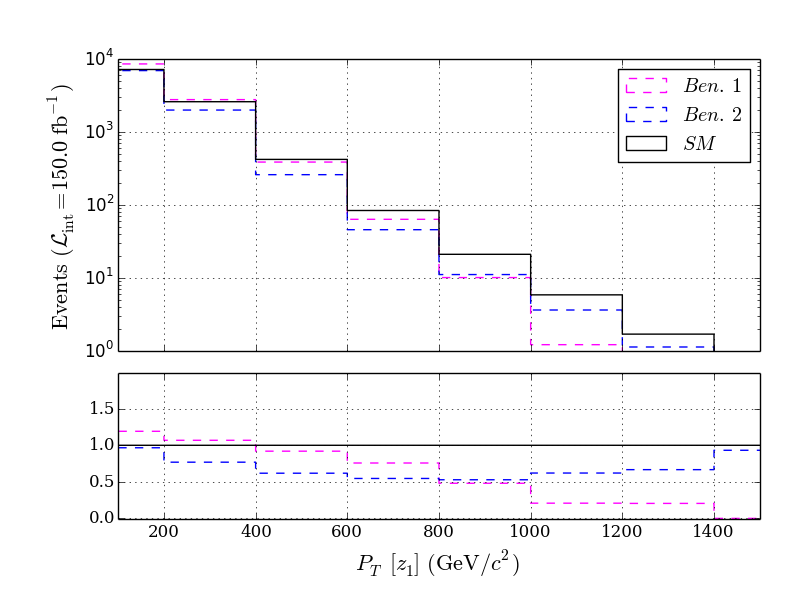}
\hfill
\includegraphics[scale=0.35]{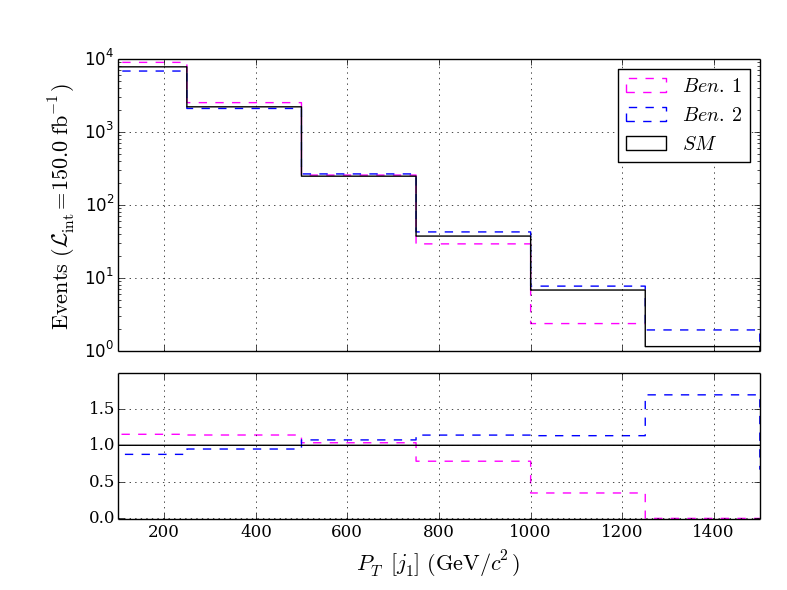}
\caption{\label{fig:bosonbenchmarks} Bosonic benchmarks B1 and B2. The ``local''  effects on individual observables and bins are very different from the global enhancement or decrease of the cross-section. The observables related with transverse momentum seem to be more discriminating than the ones related to invariant masses. As anticipated, the EFT effects are larger on the tails of the distributions. }
\end{figure}

\paragraph{Benchmarks 1 and 2} We consider all the bosonic, CP-even operators in the Warsaw basis, and given the relative contribution of each of them to the total (linear) cross section, given by equation \eqref{eq:bosonicinterf}, we find two possible solutions to the equation: one that gives an $\op(10\%)$ enhancement and one that gives a $\op(10\%)$ decrease (negative interference) to the SM total cross section. This is approximately the sensitivity we have to the process in LHC Run-2, and the order of the EW corrections to VBS at very high energies, see Refs. \cite{Denner:2000jv,Biedermann:2016yvs,Biedermann:2017bss,Biedermann:2016yds}. In Figure \ref{fig:bosonbenchmarks} we see the effects of these two benchmarks for four different VBS observables, and in table \ref{fig:bench12} we give the values used for the numerical simulation.  For this part of the study we generated $9 \cdot 10^5$ events for each benchmark scenario, using the tools described in section \ref{sec:EFTfortheGaugecouplings}.

\begin{table}[h!]
\centering
\begin{tabular}{| c ||  c | | c |}
\hline
Operator & Benchmark 1 & Benchmark 2 \\
\hline
$\bar{c}_{HB}$ &  0.0618 & -0.0157 \\ 
$\bar{c}_{H\Box}$  &  0.0620 & 0.109 \\ 
$\bar{c}_{Hd}$ &  0.0601 & 0.0872 \\ 
$\bar{c}_{HD}$ &  0.0685 & 0.0409 \\ 
$\bar{c}_{Hl}^{(3)}$ &  0.0761 & 0.0153 \\ 
$\bar{c}_{Hq}^{(1)}$ &  0.0600 & -0.0248 \\ 
$\bar{c}_{Hq}^{(3)}$ &  0.0391 & 0.0571 \\ 
$\bar{c}_{Hu}$ &  0.0601 & -0.0481 \\ 
$\bar{c}_{HW}$ &  0.0576 & -0.0360 \\ 
$\bar{c}_{HWB}$ &  0.0628 & -0.0402 \\ 
$\bar{c}_{W}$ &  0.0591 & -0.00507   \\ 
\hline
$\mu = \frac{\sigma_{EFT}}{\sigma_{SM}}$ & 0.89 & 1.14  \\
\hline
\end{tabular}
\caption{\label{fig:bench12} Benchmarks for bosonic operators. $\bar{c}_i = c_i \frac{v^2}{\Lambda^2}$. The number of significant digits used has been reduced in this table for the purpose of clarity. The CP-violating  operators $\lbrace \bar{c}_{\widetilde{W}}, \bar{c}_{\widetilde{HW}}, \bar{c}_{\widetilde{HWB}}, \bar{c}_{\widetilde{HB}} \rbrace$ have been set to zero}
\end{table}
%


\subsection{Four-fermion operators}

In this section, we repeat the same procedure for purely fermionic operators. An interesting feature of four-fermion operators is that they are always generated at tree-level in the UV-completion, whereas the rest of the dimension-six operators can be generated either at tree or loop level in the UV theory. This means that the effects of four-fermion operators will be often enhanced by a factor of $16 \pi^2$ with respect to the rest of the basis.  This is particularly  relevant for the case of study here, since purely gauge operators, those built of three field strength tensors in the \dimsix case of four field strength tensors in the \dimeight basis, are always generated from loops in the UV-completion, and hence suppressed by $16 \pi^2 = 157.914$. For the original works on the PTG/LG classification see Refs. \cite{Einhorn:2013kja}. For more recent discussions see Refs. \cite{Jenkins:2013fya, Liu:2016idz}

In this study, we consider all the fermionic operators. There are 12 operators that dominate the EFT contribution to this process, plus 3 additional ones which are very colour suppressed.  Out of these,  only $\op_{\ell \ell}$ can be constrained from the available fits, since it enters $\GF$ in the way described in section \ref{sec:notation}. The rest of the four-fermion operators remain unconstrained, to our knowledge.

\begin{figure}[h!]
\centering
\includegraphics[scale=0.35]{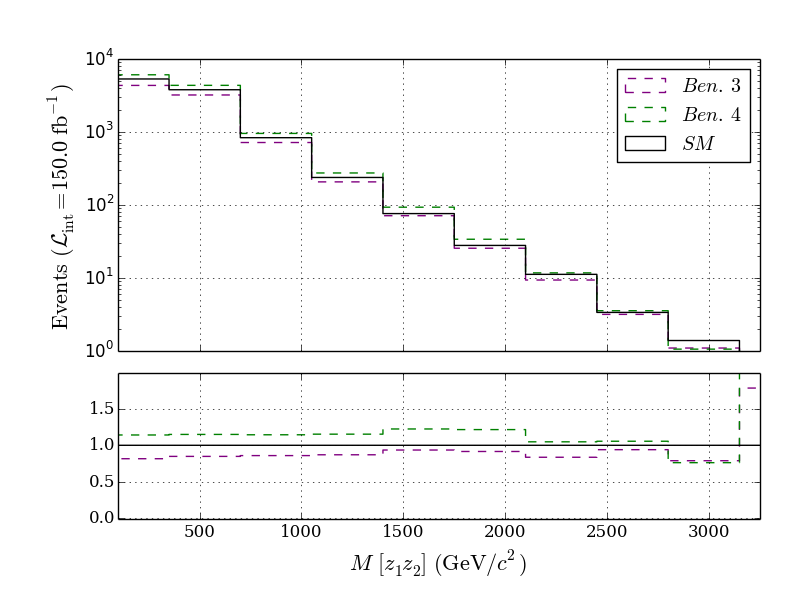}
\hfill
\includegraphics[scale=0.35]{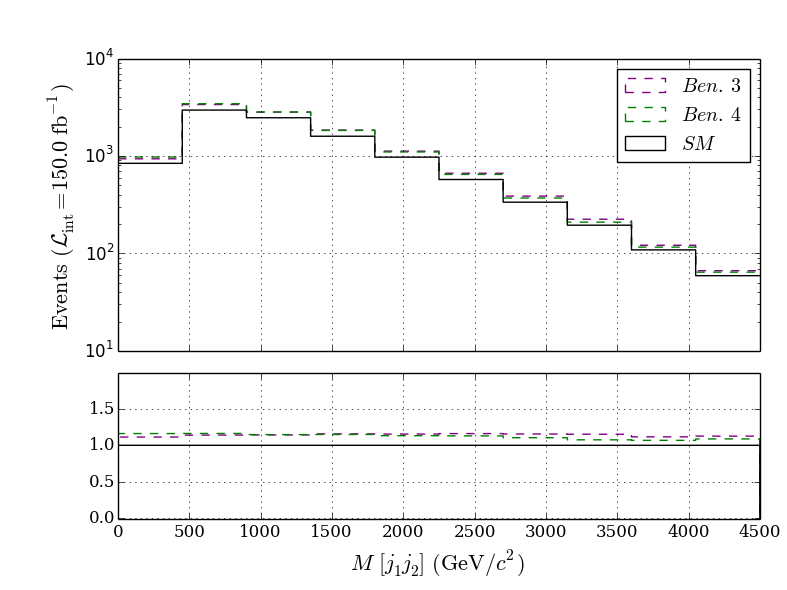}
\includegraphics[scale=0.35]{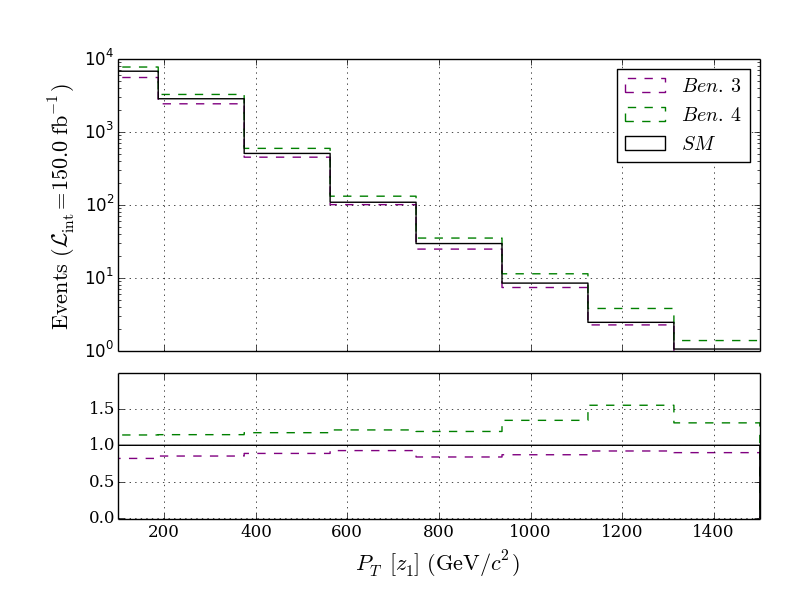}
\hfill
\includegraphics[scale=0.35]{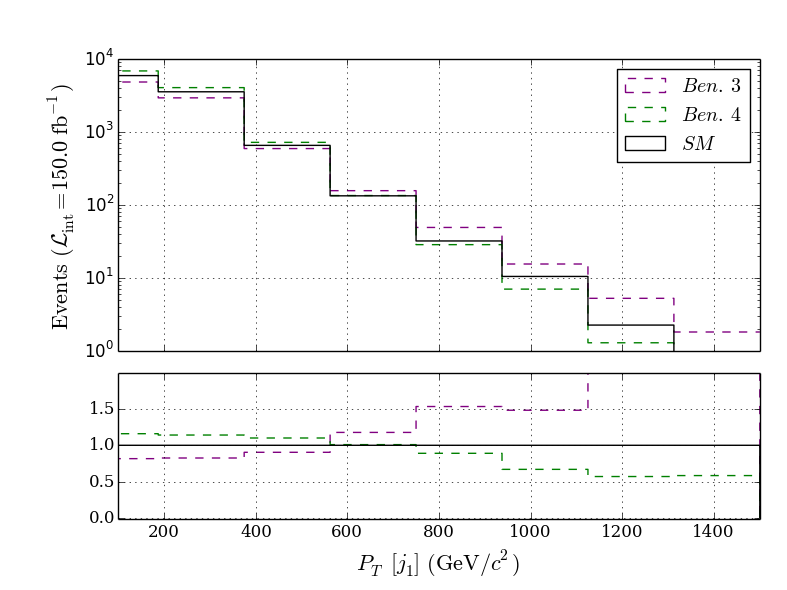}
\caption{\label{fig:fermionbenchmarks} SM prediction and benchmark scenarios from table \ref{fig:bench34}. It is interesting to see that although one benchmark gives a total enhancement to the cross-section and the other gives a decrease, in the tail of some distributions both seem to contribute positively. }
\end{figure}

\begin{figure}[h!]
\centering
\includegraphics[scale=0.35]{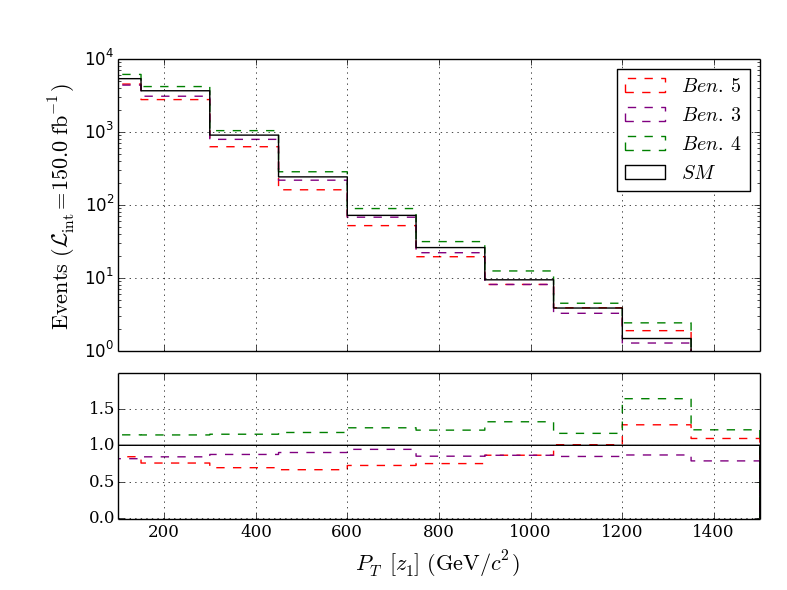}
\hfill
\includegraphics[scale=0.35]{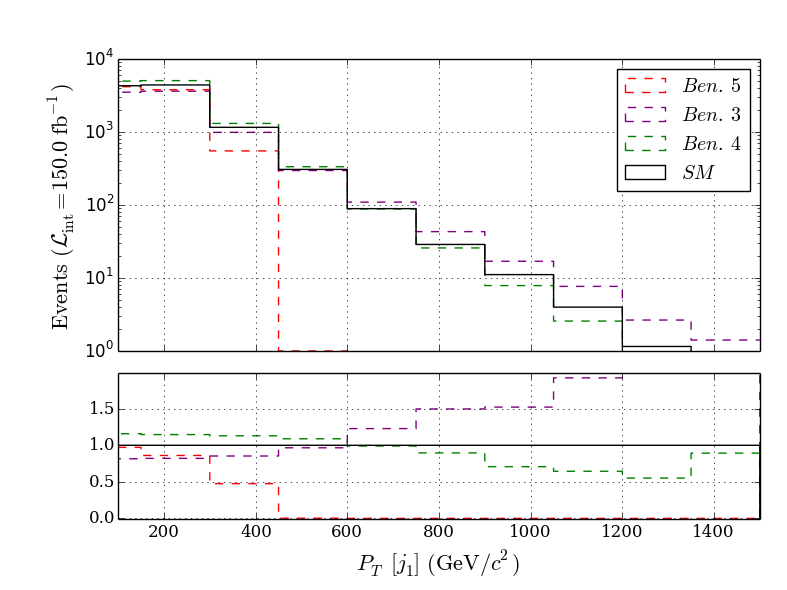}
\caption{\label{fig:fermionbenchmarks2} Example of a benchmark scenario (B5 in table \ref{fig:bench34}) that is  realistic in terms of cross section and $\pt (Z)$, but has non-physical kinematics in the  $\pt (j)$ variable. }
\end{figure}

\paragraph{Benchmarks 3, 4 and 5} Equivalently to the previous section, we find the solutions that enhance (B4) or diminish (B3) the SM cross-section by  approximately 10\%. In particular we chose the values given in table \ref{fig:bench34}. It is important to 
understand that these becnhmarks are solutions to the equation \eqref{eq:fermionicinterf}, that \emph{conspire} to enhance or decrease the SM cross-section, but the single entries of Table \ref{fig:bench34} have no physical meaning by themselves. Additionally we show the benchmark B5 which gives rise to the same total cross section as B3, but has very different kinematics.  This study can be seen in figures \ref{fig:fermionbenchmarks} and \ref{fig:fermionbenchmarks2}.

\begin{table}[h!]
\centering
\begin{tabular}{| c ||  c | | c | c | }
\hline
Operator & Benchmark 3 & Benchmark 4 & Benchmark 5 \\
\hline
$\bar{c}_{dd}$ & -0.032 &  0.061 & 0.023 \\ 
$\bar{c}_{dd}^{(1)}$  &  0.0077 &  0.061 & 0.092 \\ 
$\bar{c}_{\ell \ell}^{(1)}$ &  -0.042  &  0.039  & 0.0036 \\ 
$\bar{c}_{qd}^{(1)}$ & -0.033  & 0.060 & 0.076 \\
$\bar{c}_{qq}^{(1)}$ & 0.0099 & 0.050  &  -0.042 \\ 
$\bar{c}_{qq}^{(11)}$ & 0.024 & 0.047 & -0.043 \\ 
$\bar{c}_{qq}^{(3)}$ & 0.023 & -0.017 &  -0.042 \\
$\bar{c}_{qq}^{(31)}$ & -0.038 & 0.0031 & 0.094 \\
$\bar{c}_{qu}^{(1)}$ & 0.051 & 0.060 & -0.0020  \\ 
$\bar{c}_{ud}^{(1)}$ & 0.071 & 0.060 &  -0.084 \\
$\bar{c}_{uu}$  & -0.099  & 0.060 &  -0.037 \\
$\bar{c}_{uu}^{(1)}$ & -0.087 & 0.060 & -0.029  \\
\hline
$\mu = \frac{\sigma_{EFT}}{\sigma_{SM}}$ & 0.83 &  1.15 & 0.80  \\
\hline
\end{tabular}
\caption{\label{fig:bench34} Benchmarks for four-fermion operators. $\bar{c}_i = c_i \frac{v^2}{\Lambda^2}$. The number of significant digits used has been reduced in this table for the purpose of clarity.  We have neglected the operators containing color structures since they are very suppressed in this process, as shown in equation \eqref{eq:fermionicinterf}. }
\end{table}

 %

\clearpage
\section{Di-jet Observables}\label{sec:dijet}

Di-jet observables ($m_{jj}$, $\Delta \eta_{jj}$) are characteristic quantities to any VBS study, since the main signature of such processes is given by their jets.
 Di-jet data in general represent a great opportunity to constrain four-quark operators at LHC, some attempts in this direction have been presented in Refs. \cite{Domenech:2012ai, Alte:2017pme}. In this section we show the rapidity distributions corresponding to the different set-ups that were studied. This variable seems to be a very good pointer for NP effects in VBS.

\begin{figure}[h!]
\centering
\includegraphics[scale=0.35]{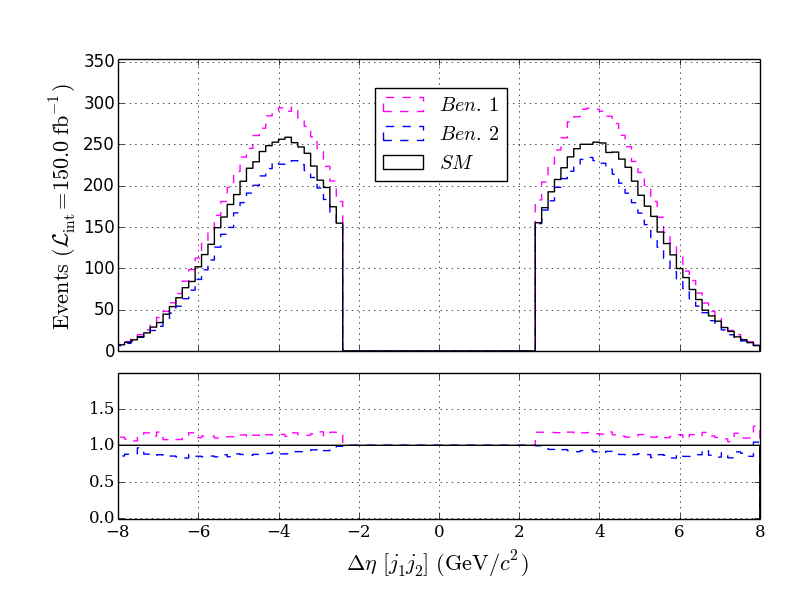}
\hfill
\includegraphics[scale=0.35]{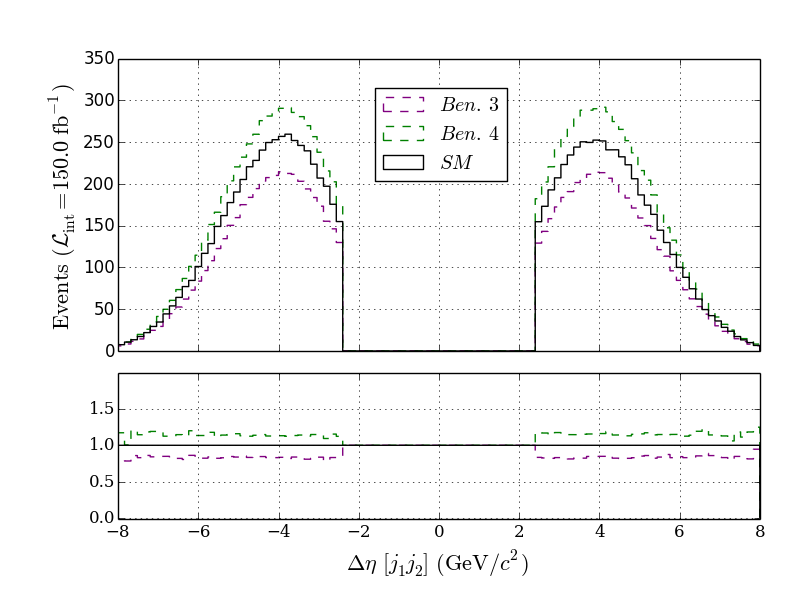}
\includegraphics[scale=0.35]{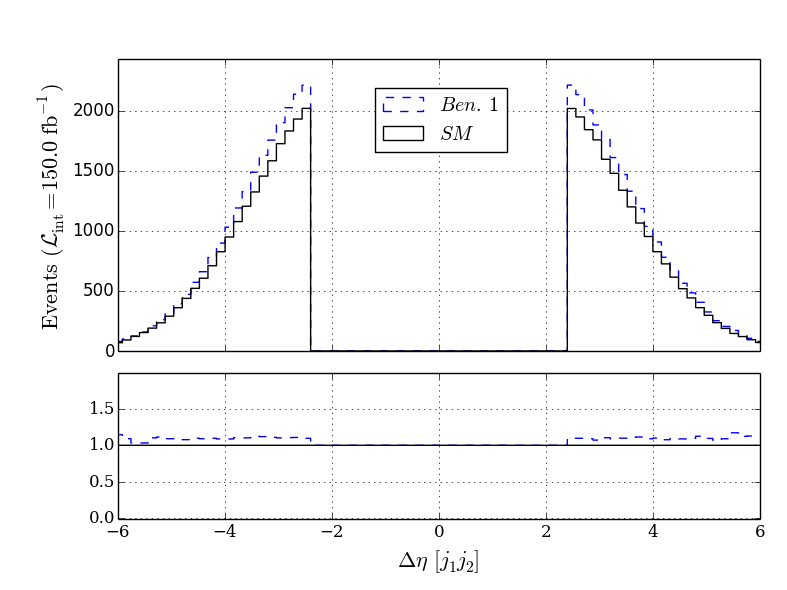}
\caption{\label{fig:bkgplots} Top: Bososnic benchmarks (left) and fermionic benchmarks (right). The $\Delta \eta$ variable seems to be a very good flag for new physics effects in VBS, where two rapidity peaks are always observed. 
Bottom: Effect of the B1 scenario in the background process, discussed in section \ref{sec:SignalAndBackground}.  The effects in the background are more subtle. Still it is important to notice that the number of background events is about one order of magnitude larger per bin, even in the VBS enriched region studied in this work. }
\end{figure}

%

\section{EFT in the Background} \label{sec:SignalAndBackground}

The main background at LHC for the previously studied process is the QCD induced $p p  \to z z j j $. Which has a very large cross-section compared with the VBS one and it is mainly discriminated from the signal thanks to the jet signature of the latter. This background was analysed in LHC in Refs. \cite{Aaboud:2017rwm,Sirunyan:2018vkx}.

Moreover, in the available VBS(ZZ) analysis, the S/B discrimination was only achieved by means of a boosted decision tree (BDT) and a Matrix Element (ME) discriminator, described in Refs. \cite{Gomez-Ambrosio:2018lcp,Sirunyan:2017fvv}.

Given this situation, it wouldn't be sensible to add the EFT effects in the signal and not in the much larger background. In the following we show a preliminary study of the EFT effects in the background and we discuss which regions and observables are best for the observation of EFT effects.

\subsection{Characterization of the background}

\begin{figure}[h]
\centering 
\includegraphics[scale=0.4]{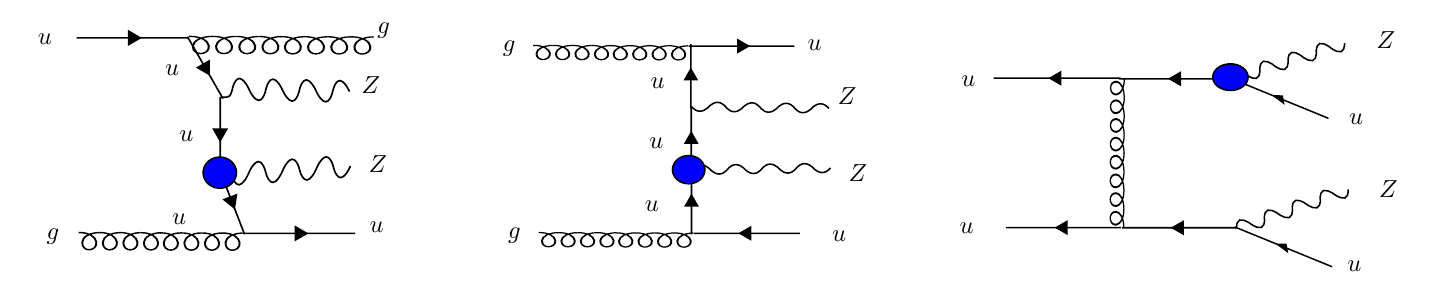}
\caption{\label{fig:bkgFD}  Some of the Feynman diagrams contributing to the VBS(ZZ) dominant background process in \dimsix EFT. The blobs represent the dimension-six insertions.}
\end{figure}

For this study, we generated a sample of events of the QCD component, i.e. $\op (g_s^2)$, to the $ p p \to Z Z j j $ process. The cross-section for this component is much larger than the purely EW one, although this discrepancy can be minimised in the VBS fiducial region, yielding:
	
\begin{equation} \label{eq:soverb}
\mu_{S/B} = \frac{\sigma_{sig.}}{\sigma_{bkg.}}\Bigg\vert_{\rm{VBS \, region}} \approx 0.25
\end{equation}
The sensitivity to the different \dimsix operators can be extracted numerically in the same way as in equations \eqref{eq:bosonicinterf}-\eqref{eq:fermionicinterf}, 

\begin{equation} \label{eq:bkginterf}
\begin{split}
\frac{\sigma_{EFT,bkg}}{\sigma_{SM,bkg}} \approx 
1. & 
- 0.00073 \, \, \bar{c}_{Hd} 
- 0.0036 \, \, \bar{c}_{HD} 
+ 0.044 \, \, \bar{c}_{HG} 
 - 0.00016 \, \, \bar{c}_{\widetilde{HG}} \\ & 
- 0.077 \, \, \bar{c}_{Hl}^{(3)} 
+  0.018 \, \, \bar{c}_{Hq}^{(1)} 
+ 0.17 \, \, \bar{c}_{Hq}^{(3)} \\ &
+  0.0065 \, \, \bar{c}_{Hu}
+ 0.035 \, \, \bar{c}_{HWB}
- 0.077  \, \, \bar{c}_{\ell \ell}^{(1)}
\end{split}
\end{equation}
This background is sensitive to less operators than the signal, and it is sensitive to other operators $\lbrace \op_{HG} , \op_{\widetilde{HG}} \rbrace$ to which the signal was blind. Some examples of Feynman diagrams for the \dimsix background can be seen in Fig.\ref{fig:bkgFD}. In order to see the effects of the \dimsix in the background, we generated a sample of background events, including the EFT operators corresponding to the benchmark scenario B1 in table \ref{fig:bench12}, and leaving the new operators equal to zero\footnote{The operator $\op_{HG}$ is very well understood through the gluon-gluon fusion process, and $\op_{\widetilde{HG}}$ is CP-odd.}. The first interesting observation is that this benchmark, that produced a negative interference of $20\%$ in the signal cross-section, gives rise to an enhancement in the background cross section, of approximately $\approx 1 \%$. Given the fact that the background cross-section is 4 times larger than the signal one, eq. \eqref{eq:soverb}, a few per-cent number of events in the background can be equivalent to an $\op(20\%)$ number of signal events.

This means that the EFT effects in the background should always be well studied and  understood. Else, an effect like this: background enhancement and signal reduction, could lead us to miss some interesting EFT effects if we only look at signals and total rates. 

In figure \ref{fig:bkgplots} we show the corresponding plots for the background study. 

\begin{figure}[h!]
\centering
\includegraphics[scale=0.35]{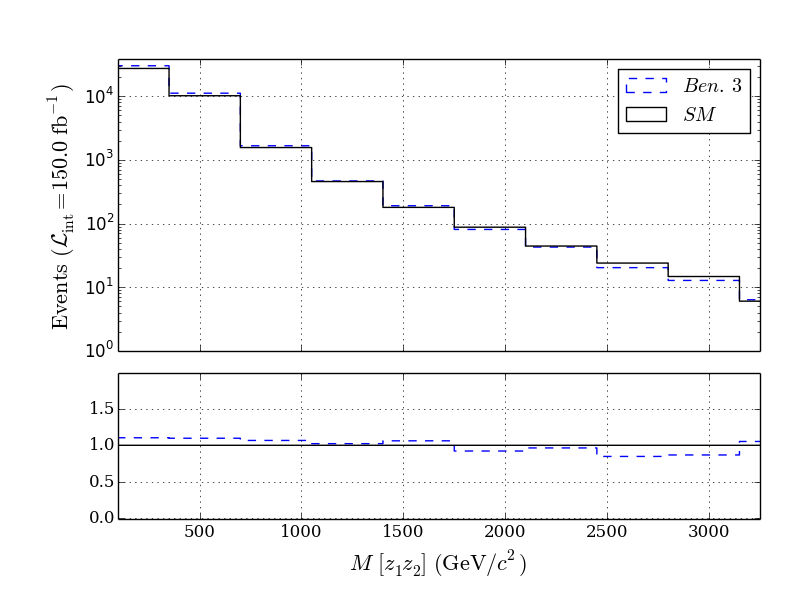}
\hfill
\includegraphics[scale=0.35]{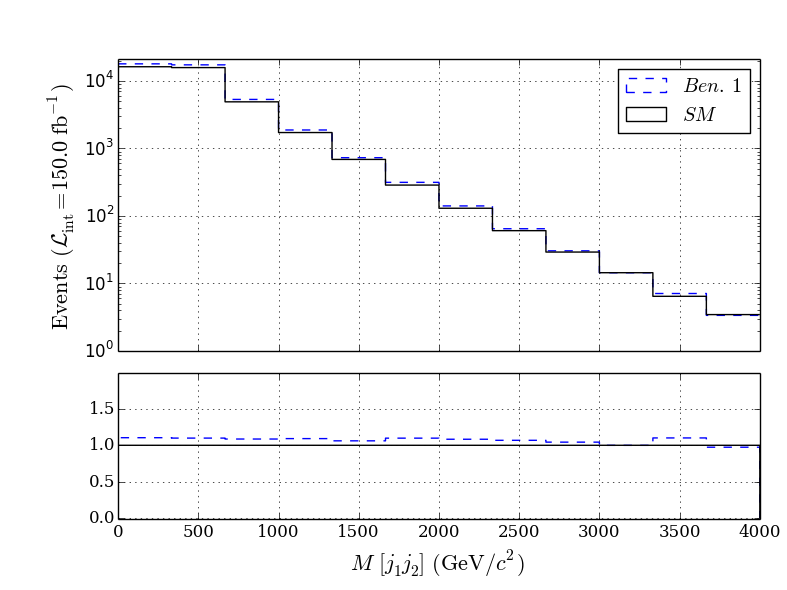}
\includegraphics[scale=0.35]{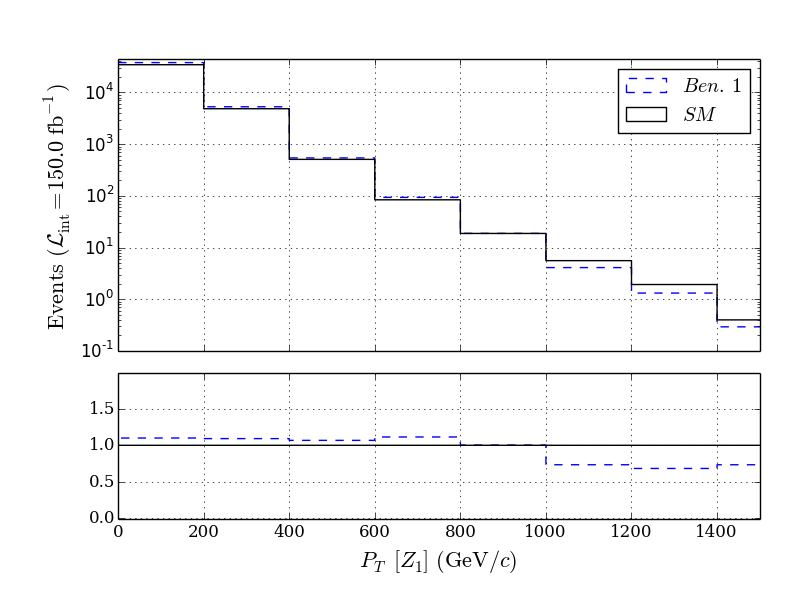}
\hfill
\includegraphics[scale=0.35]{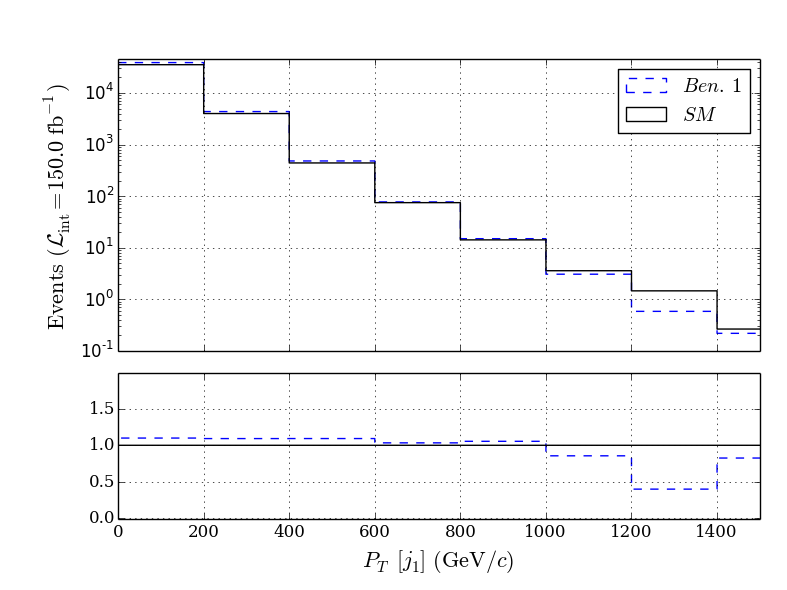}
\caption{\label{fig:bkgplots} Effect of the B1 scenario in the background process. The effect is in principle very small, however it is important to remark that the number of background events is about one order of magnitude larger per bin, than that for the signal. }
\end{figure}

%

%
%
%

%
%
%
%


 %

\clearpage
\section{General Strategy for EFT studies at LHC} \label{sec:generalStrategy}
 
In this work we have shown that at least 11 bosonic and 12 fermionic dimension-six operators enter the VBS(ZZ) signal cross-section. If we try to isolate the effects of each of these operators using a single cross-section measurement, the solutions to this system define a 22 dimensional manifold. In order to  further constrain the space of solutions it is necessary to add further experimental measurements. 

In this work we have proposed different VBS observables\footnote{We focus on the ZZ channel here, but his logic applies equally to the other VBS channels.}: invariant mass of the two gauge bosons, invariant mass and rapidity of the di-jet system, transverse momentum of the leading gauge boson and jet and the rapidity separation between jets. As we have seen throughout section \ref{sec:warsawVBS}, the EFT scales differently in different regions of phase space (with the high energy ``tails'' being privileged). This means that a differential measurement with $n$-bins could be used to add $n$-equations to the aforementioned 23-variable system, reducing the number of directions in the solution-hyperplane. 

Moreover, if we assume \emph{factorisation} of the EFT effects, by defining the amplitude as in section \ref{sec:amplitudes}, with one insertion per diagram and at tree-level, the fiducial cross-section can be written as,
\begin{equation}
\sigma_{EFT} = \sigma_{SM} + A \cdot (
\mathcal{F} ( \bar{c}_1 , \bar{c}_2 , \dots , \bar{c}_{23} )
), 
\end{equation}
where $\mathcal{F}$ is a linear function of the Wilson coefficients $\bar{c}_i$, and $A$ is a global factor modulating the EFT contribution. In that case, the most important task is to find the relationship between the coefficients, $\mathcal{F}$, since the global factor $A$ can be constrained with the measurement of the fiducial cross-section. The best way to do that includes using multivariate analysis (MVA) techniques, like the ones currently used in the experimental VBS searches. For example in the analysis discussed here, or in the analysis of the VBS(WZ) in ATLAS, in Ref.\cite{ATLAS:2018ucv}, where a 15-variable BDT was used.

It is always possible to add different measurements (from LEP, Higgs physics, \emph{etc.}), in order to reduce the number of unknowns in the system of equations. But this should be dealt-with with great care: a bound set for a certain Wilson coefficient is only valid in the energy regime where the calculation is performed. In order to know the value of this coefficient at a different energy scale, the renormalization group evolution of the coefficient has to be calculated. In that regard, it is safer to include more observables and more bins for signal and background measurements than to mix VBS-signal, Higgs-signal and LEP-signal measurements. 

Ideally, gauge boson polarization measurements, not studied in detail here, could also be useful. Such polarization studies happen to be very interesting already at the SM level,as pointed out in recent works like Ref. \cite{Ballestrero:2017bxn,Brass:2018hfw}. Further, the study of the parton shower effects on the EFT distributions, might also shed new light on the problem. There are very few references in this direction so far, for example Ref. \cite{Alioli:2018ljm}. Last but not least, the study of the backgrounds, as shown in section \ref{sec:SignalAndBackground}, is also necessary for a correct determination of the EFT effects.

\section{Conclusions}

At the current level of precision of LHC measurements, and the high level of agreement of the former with the SM predictions, it is advisable to perform any EFT analysis with a great deal of accuracy. 
For this purpose EFT operators cannot be studied on a case-by-case basis, and a global study of the set of dimension-six operators is necessary in the first place. In a second stage, dimension-eight as well as NLO and quadratic dimension-six effects need to  be studied in order to improve the EFT theoretical uncertainties. 


\acknowledgments

We want to thank Giampiero Passarino for relevant comments on the manuscript, and Roberto Covarelli for relevant comments regarding the experimental set-up. We gratefully acknowledge Michael Trott for providing us with the likelihood functions for the fits presented in Refs.\cite{Berthier:2016tkq,Brivio:2017bnu} as well as Ilaria Brivio for technical support with the SMEFTsim package, very interesting comments and discussions, and relevant input at the beginning of this project. We also want to thank Pietro Govoni for hosting us at the initial stage of this project, supported by a STSM Grant from COST Action CA16108. As well as the other members of the Action for interesting discussions.

\bibliographystyle{atlasnote}
\bibliography{thesisBib}

\end{document}